\documentstyle[12pt,aaspp4]{article}

\def\t0{\theta_{\circ}}

\def\be{\begin{equation}}
\def\en{\end{equation}}

\def\msun{M_{\sun}}

\def\lsun{L_{\sun}}
\def\msunyr{M_{\sun} yr^{-1}}
\def\kms{\rm \, km \, s^{-1}}
\def\md{\dot{M}}
\def\Md{\dot{M}}

\begin{document}

\title {Infall models of Class 0 protostars}
\author{Ray Jayawardhana}
\affil{Department of Astronomy, University of California, Berkeley, CA 94720}
\author{Lee Hartmann and Nuria Calvet}
\affil{Harvard-Smithsonian Center for Astrophysics, 60 Garden Street, 
Cambridge, MA 02138; email: lhartmann@cfa.harvard.edu}

\begin{abstract}
We have carried out radiative transfer calculations of infalling, dusty 
envelopes surrounding embedded protostars to understand the observed 
properties of the recently identified ``Class 0'' sources. To match the 
far-infrared peaks in the spectral energy distributions
of objects such as the prototype Class 0 source VLA 1623, pure 
collapse models require mass infall rates $\sim$10$^{-4}\msun$yr$^{-1}$. 
The radial intensity distributions predicted by such infall
models are inconsistent with observations of VLA 1623 at sub-mm wavelengths, 
in agreement with the results of Andr$\acute{\rm e}$ et al. (1993) who found 
a density profile of $\rho \propto r^{-1/2}$ rather than the 
expected $\rho \propto r^{-3/2}$ gradient.
To resolve this conflict, while still invoking infall to produce the outflow
source at the center of VLA 1623, we suggest that the observed sub-mm 
intensity distribution is the sum of two components: an inner infall zone, 
plus an outer, more nearly constant-density region.  This explanation of the 
observations requires that roughly half the total mass observed
within 2000 AU radius of the source lies in a region external to
the infall zone.  The column densities for this external
region are comparable to those found in the larger Oph A cloud 
within which VLA 1623 is embedded.  This decomposition into infall 
and external regions is not unique, owing to uncertainty in the structure 
of the molecular gas outside of the infall zone, which in turn implies
some uncertainty in estimating the infall rate.  Nevertheless,
the environment of Oph A is so dense that any protostellar clouds 
which fragment out are likely to collapse at very high infall rates, consistent
with our spectral energy distribution modeling. The extreme environments 
of Class 0 sources lead us to suggest an alternative or
additional interpretation of these objects: rather than simply concluding with 
Andr$\acute{\rm e}$ et al. that Class 0 objects only represent the earliest 
phases of protostellar collapse, and ultimately evolve
into older ``Class I'' protostars, we suggest that 
many Class 0 sources could be the protostars of very dense regions, 
and Class I objects found in lower-density regions 
may be in comparable evolutionary states.

\end{abstract}


\newpage
\section{Introduction}
An important aim of the study of young stellar objects (YSOs) has
been to classify them along an evolutionary sequence. According to the
widely used classification scheme of Lada \& Wilking (1984) 
and Lada (1987), YSOs are divided into three classes
based on the slopes of their infrared spectral energy distribution (SED).
The reddest objects, dubbed Class I sources,
are thought to be protostars surrounded by infalling envelopes. Their
SEDs are consistent with dust emission from envelopes with infall
rates near those predicted by the standard theory of protostellar formation, 
which invokes free-fall collapse from an initially thermally-supported, 
isothermal, singular molecular cloud core in hydrostatic equilibrium (Adams 
\& Shu 1986; Adams, Lada \& Shu 1987). Kenyon, Calvet \& Hartmann (1993, 
hereafter KCH), in particular, were able to match the observed SEDs of
Class I sources in Taurus-Auriga with rotating infall models of the type
described by Terebey, Shu \& Cassen (1984). This agreement between theory
and observation is generally seen as testimony to the success of the
standard model of protostellar collapse.

Recently, Andr$\acute{\rm e}$, Ward-Thompson \& Barsony (1993, hereafter
AWB) identified YSOs characterized by blackbody-like SEDs 
that peak at submillimeter wavelengths. The low dust temperatures 
($T_{d}$ $\sim$ 20K) and large dust masses inferred for
these sources led AWB to suggest that these objects constitute a new
class of objects, called ``Class 0'', which might be even younger than 
typical Class I objects. Observationally, the defining characteristics of 
Class 0 sources include a high ratio of submillimeter to bolometric 
luminosity (L$_{submm}^{\lambda > 350\mu m}$/L$_{bol}$ $>$ 5
$\times$ 10$^{-3}$), invisibility at near-infrared wavelengths, and the 
presence of molecular outflows (Barsony 1994). Andr$\acute{\rm e}$ \& 
Montmerle (1994) argued that Class 0 sources have more circumstellar 
mass than stellar mass. In this picture, Class 0 sources have not yet 
assembled the bulk of their final stellar mass and are still in the main 
accretion phase. Consequently, the authors argue that Class 0 sources are 
the youngest protostars, and subsequently evolve into Class I sources
(see Andr\'e, Ward-Thompson, \& Barsony 2000 $=$ AWB2).

Following AWB's identification of the prototype
Class 0 source VLA 1623 in the rich star-forming core $\rho$ Ophiuchi A,
several other objects of the class have been found in Orion B 
(Ward-Thompson et al. 1995), Serpens (Hurt \& Barsony 1996), Cep E (Lefloch 
et al. 1996), and Perseus (O'Linger et al. 1997) molecular clouds. In a 
spectroscopic survey of 47 candidate protostars, Mardones et al. (1997) 
report that Class 0 sources have stronger infall signatures --i.e., larger 
negative velocity shifts-- than Class I sources, and attribute that result
to a difference in the physical conditions in the circumstellar envelopes 
of the two source types.

In order to understand better the observed characteristics of this `youngest'
population of YSOs, we have performed detailed radiative transfer 
calculations of dust envelope emission in Class 0 sources, following 
the methods described by KCH and Calvet et al. (1994). We also study 
the observed radial intensity profiles of the spatially-resolved prototype
Class 0 source VLA 1623 at three submillimeter wavelengths.
In both cases, we compare the observations to the predictions
of the standard model, and explore the possibility that environments exterior 
to the infall region can affect the spectral energy distribution and the 
inferred density gradients.

\section{Standard Model}
\subsection{Methods and Parameters}
Following KCH, we adopt the Terebey, Shu \& Cassen (1984, hereafter TSC)
solution for a uniformly rotating protostellar cloud with a radial density
distribution initially approaching that of a singular isothermal sphere, 
$\rho(r) \propto r^{-2}$. However, unlike TSC, we assume an outer boundary 
to the envelope, $r_{out}$. Inside the collapse region, the infall velocity 
approaches the free fall value, $v \propto r^{-1/2}$, and the density 
distribution converges toward $\rho \propto r^{-3/2}$, because the mass 
infall rate, $\Md_{i}$ is constant. Inside the centrifugal radius,
$R_{c}$, material falls onto a circumstellar disk rather than radially onto
the central object.  

We use the {\it angle-averaged} density distribution to construct the spherically
symmetric source function, $S_{\nu}=\sigma_{\nu}J_{\nu}+(1-\sigma_{\nu})B_{\nu}$, 
where $\sigma_{\nu}$ is the albedo and $B_{\nu}$ is the Planck function. Scattering
increases the flux at short wavelengths, $\lambda \lesssim 3 \mu$m. We then derive
the emergent flux from the envelope using $S_{\nu}$ for the source function
and the {\it angle-dependent} density, $\rho(r,\theta)$, for the opacity. 

The angular momentum of the infalling material modifies the density 
distribution from the purely-radial infall solution at $r \lesssim R_c$.  
However, for the models considered here, the observable emission originates 
from envelope radii much larger than typical disk sizes $\sim 100$ AU, and 
so the value of $R_c$ is unimportant.  Similarly, we do not include envelope 
cavities such as those produced by bipolar outflows.  KCH pointed out that 
envelope cavities were needed to explain the near-infrared emission of some 
Taurus Class I sources; but the very large optical depths characteristic of 
the models considered here prevent the escape of any significant amount of 
near-infrared scattered light from the envelope.

Thus, our models consist of a central stellar source, with bolometric luminosity
\emph {L}, surrounded by an envelope that is essentially spherically-symmetric, 
with inner radius $r_{in}$ and outer radius $r_{out}$. The inner radius of the 
envelope is set by the dust destruction radius, which is assumed to occur at $T \sim$ 
1500K. For $r_{out}$, we adopt values roughly equivalent to the observed sizes of the 
sources being modeled.  The models have then three free parameters: the source 
luminosity \emph{L}, $r_{out}$, and the density of the envelope (see below). We used 
Draine \& Lee (1984) opacities for $\lambda \le 200 \mu m$ and a shallower 
$\beta =$1.5 (where $\kappa_{\nu} \propto \nu^{\beta}$) opacity law for longer 
wavelengths.

\subsection{Model Results}
\subsubsection{Spectral energy distribution}
We began by constructing a grid of models with a wide range of envelope
densities for source luminosities consistent with the observed values
for Class 0 sources of interest. These models were defined in terms of $\rho_{1}$, 
the density at \emph{r}= 1 AU in the limit that $R_{c} \rightarrow$ 0. The mass 
infall rate is related to this density parameter and the central mass by 
\begin{equation}
\Md = 1.9 \times 10^{-6} \msun \: \mathrm{yr^{-1}} \left (\frac{\rho_{1}}{10^{-14} \mathrm{g \: cm^{-3}}} \right ) \left ( \frac{\mathit{M_{*}}}{1 \mathit{\msun}} \right )^{1/2}
\end{equation}
The envelope temperature distribution is not very sensitive to the SED of the central
source because its radiation is immediately absorbed by the innermost
(i.e., large optical depth) regions of the envelope and reradiated at
longer wavelengths. 

These models show general agreement with the analytic approximation of
KCH for the peak wavelength of the SED:
\begin{equation}
\lambda_{m} \approx 60 \left (\frac{L}{10^{33} \mathrm {\: erg \: s^{-1}}} \right )^{-1/12} \left (\frac{\rho_{1}}{10^{-13} \mathrm {\: g \: cm^{-3}}} \right )^{1/3} \: \mathrm{\mu m.}
\end{equation}
For typical source luminosities and expected central masses,
in order to shift the peak of the SED to $\lambda \sim 200 \mu m$, 
as is observed in Class 0 sources, the models require infall rates 
$\geq 10^{-4} \msun yr^{-1}$. 

Unfortunately, SEDs of most known Class 0 sources are not well defined
observationally. Many of them have been observed only at a couple of 
sub-millimeter wavelengths, and the IRAS upper limits often do not provide 
strong constraints in the mid- and far-infrared. Therefore, we restrict our
detailed modeling to two sources with relatively well-defined SEDs:

{\it VLA 1623} Located in the $\rho$ Oph A molecular cloud, VLA 1623
is a centrally-peaked source of millimeter emission (Andr$\acute{\rm e}$
et al. 1990b), a weak 6 cm radio source (Leous et al. 1991), and the driving
source of a strong CO bipolar outflow (Andr$\acute{\rm e}$ et al. 1990a).
The highly collimated outflow has a mechanical luminosity of 
$L_{flow} \sim 0.5-2 \lsun$ and a short dynamical timescale 
$t_{D} \sim 3000-6000$ yrs, making it one of the youngest known flows.  
VLA 1623 is undetected at near- and mid-infrared wavelengths to very low
limits. Based on these properties as well as its remarkably cold temperature 
($T \sim$ 20K), high internal obscuration, apparently massive circumstellar 
structure and high $L_{submm}/L_{bol}$ ratio, AWB 
proposed it as the prototype of the newly-defined Class 0 sources.
Observations with the JCMT-CSO interferometer show an unresolved central
source --presumably a circumstellar disk-- in VLA 1623. However, its
contribution to the total flux at 1.3mm and 800$\mu$m is $\leq$ 10\% 
(Pudritz et al. 1996). 

In our infall model for VLA 1623, we set $L=1\lsun$ and $r_{out}$=2000AU,
which corresponds to 12'' at the distance of Ophiuchus and is roughly the 
observed source size in the sub-millimeter. (We 
adopt $d=160$ pc to Oph, the same as AWB.) For the model 
that best fits the observed SED of VLA 1623 (Fig. 1a), the mass infall rate is
$\sim 2 \times 10^{-4} \msun \mathrm{yr^{-1}}$ for a central stellar mass
of 0.5$\msun$. A lower infall rate model, with $\Md \sim 10^{-5}
\msun \mathrm{yr^{-1}}$, shown as a dotted line in Fig. 1a, does not match
the data. Using the same parameterization of opacity as AWB --
$\kappa_{\nu} = 0.1 \mathrm{cm^{2}g^{-1}} (\nu/1000 GHz)^{1.5}$-- we also 
obtain a total envelope mass of $\sim 0.6 \msun$ for the best-fit model.

{\it HH24MMS} This bright source of millimeter emission near the Herbig-Haro
object HH24 in the Orion B molecular cloud at a distance of 460 pc was 
discovered by Chini et al. (1993). It is not detected in the near- or 
far-infrared, although IRAS limits are rather poor because of source 
confusion. VLA observations have revealed a 3.6cm source coincident with 
the millimeter peak (Bontemps, Andr$\acute{\rm e}$ \& Ward-Thompson 1995). 
Using JCMT observations at 350, 450, 800 and 1100$\mu$m, Ward-Thompson et 
al. (1995) have produced a reasonably good SED for HH24MMS, even though its 
peak is not tightly constrained due to a lack of shorter wavelength 
detections.

We modeled HH24MMS with $L=2.5\lsun$ and $r_{out}$=4000AU, corresponding
to an angular cloud diameter of 18''. Our best fit model requires a mass 
infall rate of $\sim 3 \times 10^{-4}\msun \mathrm{yr^{-1}}$, assuming
$M_{*}=0.5 \msun$ (Fig. 1b). 

\subsubsection{Radial intensity profiles}
Since VLA 1623 is slightly extended at sub-millimeter wavelengths, we
attempted to compare our ``best-fit'' model profiles to the observed
azimuthally-averaged radial intensity profiles at 350$\mu$m, 450$\mu$m and
800$\mu$m of AWB. The model output was convolved with the appropriate beam 
size at each wavelength and scaled to a distance of 160 pc. All the profiles, 
shown as dotted lines in Figure 2, have been normalized to the peak flux 
values. As Andr$\acute{\rm e}$ et al. noted, a modified TSC infall model with 
$\rho \propto r^{-3/2}$ cannot reproduce the observed profiles, which are 
significantly shallower. 

In our basic model, the temperature at the outer edge of the infalling cloud
is not constrained and in fact drops to $\sim$ 7 K. Given that external
heating is likely to keep the cloud surface at $\sim$ 20K, we explored 
the effect of adopting a minimum temperature in the envelope
of $T_{min} =$ 20K. The result,
shown as dashed lines in Figure 2, is to produce somewhat shallower
intensity distributions, but still not shallow enough to fit the observations
well.

Our calculations differ from those of Terebey, Chandler \& Andr\'e (1993)
in that we assume an outer boundary to the envelope. The inability to 
determine the zero flux level in the observed intensity profiles is an
additional source of uncertainty in our fits.

\section{Contribution of exterior regions}
The presence of a compact VLA source and a collimated CO bipolar outflow
as well as the centrally peaked submillimeter emission strongly suggest
that a central source --i.e., a protostar-- has already formed in 
VLA 1623.  Moreover, the presence of a compact ($< 70$~AU) mm and submm
source consistent with a dusty circumstellar disk
(Pudritz et al. 1996) emphasizes the likelihood
that substantial infall has already occurred.
Thus, one would expect to find an infall region characterized
by a density distribution of roughly $\rho \propto r^{-3/2}$ (see \S 4.1). 
In this context the $\rho \propto r^{-1/2}$ distribution inferred 
by AWB from the radial intensity profiles is surprising. 
\footnote{A $\rho \propto r^{-1/2}$ density distribution
corresponds to $\Sigma \propto r^{1/2}$ surface density distribution.  
Thus, for the slowly-varying temperature distributions characteristic
of the outer envelope $T \approx r^{-0.4}$, the surface brightness distribution
would actually increase with increasing radius without including an
outer cutoff radius; this parameter has a crucial effect on the resulting
intensity distribution.}

One possibility is that material outside the infall region contributes to
the observed intensity profile. As clearly seen in the submillimeter 
continuum maps of $\rho$ Ophiuchi (AWB; Motte et al. 1998), VLA 1623 is 
indeed surrounded by --and likely embedded in-- larger-scale cloud emission. 
We have attempted to model the exterior cloud material as a uniformly
emitting ``slab'' at 20K. The resulting radial intensity 
profiles, which include contributions from the infall zone as well as the 
exterior region, are shown as solid lines in Figure 2. We find that in order 
to match the observed intensity distributions at 350$\mu$m, 450$\mu$m and 
800$\mu$m, the ``slab'' needs to have a column density 
$< N_{H2} > \sim 6 \times 10^{22}$ cm$^{-2}$ or $A_{V} \approx 60$, 
implying $M_{slab} \approx 0.3 \msun$ in the 2000 AU radius region. 
This compares well with typical column densities of $\approx 10^{23}$ in the 
larger $\rho$ Oph A cloud derived by Motte et al. (1998). 

Could extinction by this outer region make a Class I source appear as
a Class 0? A ``slab'' of the kind we propose, with $A_{V} \approx 60$,
would significantly extinct emission from the inner regions at short wavelengths.
However, because the extinction of such an outer region is
$A_{\lambda} < 1$ at wavelengths $\lambda \gtrsim 30 \mu m$, it would
not shift the peak of a Class I SED to $\lambda > 100 \mu m$. Therefore, it 
is unlikely that a Class I source with a relatively flat SED in 
$\lambda$ F$_{\lambda}$ would appear as a Class 0 simply due to external extinction.

Our best-fit, composite model for the SED of VLA 1623 is shown in
Figure 3. Using the same opacities as AWB, this object has a
total mass of $0.6 \msun$, divided roughly equally between the central
infall zone and the outer, nearly constant-density region. Our composite 
model still requires a high infall rate of $\sim 10^{-4}\msun 
\mathrm{yr^{-1}} (M_*/0.5 \msun)^{1/2}$, 
i.e., the necessary infall rate is only reduced by a 
factor of two by the inclusion of the constant density outer region.

Interestingly, model fits to the `youngest' (i.e., most embedded) 
protostars in Chandler \& Richer's (2000) sample imply that the power-law
index of the envelope density distribution $p$ (defined by $\rho \propto
r^{-p}$) is lower than that predicted by the standard collapse model.
They suggest that a shallow density profile may be maintained in the outer
envelope by magnetic fields and/or turbulence. On the other hand, 
Hogerheijde \& Sandell (2000) find that the density distribution in 
envelopes around four embedded YSOs in Taurus are fit equally well by 
a radial power-law index $p=$1.0--2.0 or by a Shu collapse model. 

\section{Discussion}
\subsection{Environment and high infall rates}

In the previous section we showed that models of a collapsing isolated
cloud can fit the SEDs of Class 0 source if a high infall rate is adopted,
but that the resulting surface brightness distributions are inconsistent with
observations, as originally found by AWB.  We have extended this result
to include the effects of external heating, and still find a discrepancy 
between
theory and observation.  The surface brightness distributions can be explained
with simple infall models only by including an additional emitting component, 
which could be material in the near environment of the collapsing cloud.

Infall models generally exhibit surface density distributions 
$\Sigma \propto r^{-1/2}$ (implying $\rho \propto r^{-3/2}$ 
in spherical geometry), whether from spherical or flattened clouds, 
similarity solutions for collapse or numerical time-dependent calculations,
with or without cloud magnetization (e.g., Larson 1969; Penston 1969; 
Shu 1977; Foster \& Chevalier 1993; Galli \& Shu 1993a,b; Hartmann et al. 
1994; Hartmann, Calvet, \& Boss 1996; Li \& Shu 1996).
In isothermal clouds at the point of collapse, many calculations
show density distributions which tend to steepen more towards $r^{-2}$ 
(e.g., Mouschovias 1991, and references therein; Foster \& Chevalier 1993).  
These models strongly suggest that the observations of Class 0 sources
encompass not only an infall region, but external material not necessarily
in collapse.

The question then arises: if the external environment contributes a 
large fraction of the sub-mm flux, can we in fact say anything about infall 
rates?  A static cloud, with density distribution
$\rho \sim r^{-1/2}$ as considered by AWB, could, if physically realizable,
fit the observations in the absence of {\it any} infall.
(Note that such a static cloud is inconsistent with the standard model
of isothermal clouds, in which a $\rho \propto r^{-2}$ density distribution
is expected). It seems likely that some fraction of the sub-mm emission
arises from an infall region, because some infall must have occurred to produce
the central source, and given that pre-stellar cores appear to be
much less centrally condensed than Class 0 sources (Ward-Thompson et al. 1994;
Andr\'e et al. 1996; Ward-Thompson, Motte, \& Andr\'e 1999; AWB2). 

Nevertheless, let us take the extreme limit by supposing
that {\it all} of the sub-mm emission arises from a static cloud. 
We may then estimate the rate at which this cloud would collapse.
For a cloud mass of $\sim 0.6 \msun$ concentrated within a radius of 2000 AU,
the free-fall timescale is $\sim 2 \times 10^{4}$ years. Thus, if this cloud 
were to collapse, it would exhibit an average infall rate of 
$\sim 3 \times 10^{-5} \msunyr$. 
This estimated infall rate is not much smaller than the $\sim 10^{-4} \msunyr$
estimated for the pure infalling envelope model based on the SED fitting.
Moreover, if the region that fell in to form the outflow source
was originally {\it denser} than the outer static region, as seems probable, 
the infall rate could be higher than this estimate.
Thus, one is lead to the same order of magnitude infall rate from consideration 
of the spatial concentration of the local gas, independent of the fraction of emission 
currently arising from an external cloud rather than the infall region.

High infall rates are also suggested by observations of outflows.
In a survey of nearby embedded YSOs, Bontemps et al. (1996) found that
Class 0 sources have unusually powerful outflows. The Class 0 objects
in their sample lie an order of magnitude above the well-known correlation
between the outflow momentum flux (integrated over
time) and bolometric luminosity that holds for
Class I sources. Since models of bipolar outflows (Shu et al. 1994, 
Ferreira \& Pelletier 1995) predict that the mass-loss rate is some
fraction of the mass-accretion rate, the stronger outflows in Class 0s
also call for significantly higher infall rates than in Class Is.
Henriksen, Andr\'e, \& Bontemps (1997) attempt to derive accretion rates 
as a function of time for Class 0 sources which are consistent with infall 
at non-constant rates, again finding very rapid mass accumulation is required 
to explain outflow properties.

In addition, from a spectroscopic survey of 47 candidate protostars, 
Mardones et al. (1997) report that Class 0 sources have stronger infall 
signatures --i.e., larger negative velocity shifts-- than Class I sources.
One possible interpretation of their results is that Class 0s have 
higher density envelopes than Class Is, again implying higher infall
rates in Class 0 objects.

Thus, our high infall rates, which imply that Class 0 sources are a short-lived 
phenomenon, are consistent with previous conclusions in the literature 
(e.g., AWB). A comparison of source statistics in Taurus-Auriga and Ophiuchus 
support this conclusion. With an infall rate of $10^{-4} \mathrm{\msunyr}$, a
0.5 $\msun$ star would be accreted in $\sim$5000 years, a timescale consistent
with the dynamical lifetime of VLA 1623.  Andr\'e \& Montmerle (1994) and Barsony (1994)
estimate a lifetime of $\sim 10^4$ yr for Class 0 sources in Oph on statistical grounds. 
In addition, as there are around 200 
known pre-main-sequence objects in Taurus, one would expect 
5000 yrs$/10^{6}$ yrs $\times$ 200 or $\sim$1 Class 0 source; and 
Taurus-Auriga contains only two ``border-line'' Class 0 sources,
L1527 and L1551 IRS 5 (Chen et al. 1995).
(AWB2 suggest that two Class 0 sources are present in Taurus, L1527 
and the extremely low-luminosity source IRAM 04191+1522 [Andr\'e, Motte, \& Bacmann 1999]).

One problem with the picture of high infall rates for Class 0 sources
is that much higher accretion rates onto the central protostar should lead
to high accretion luminosities. However, Class 0 objects
are not very much more luminous than their Class I counterparts. Henriksen,
Andr\'e, \& Bontemps (1997) and Andr$\acute{\rm e}$ (1997) have suggested
several factors that may reduce the theoretical accretion luminosity: in 
Class 0 sources, the central stellar mass might be lower, the stellar radius 
probably larger, and the amount of accretion energy dissipated in the wind 
is also likely to be larger. Another natural explanation is that material
falls onto a disk, instead of falling directly onto the central star. 
However, the material must still somehow flow from the disk to the star
at some point in the protostellar phase to build up the central star. It 
may be that Class 0 sources also undergo ``FU Orionis-type'' outbursts once 
enough material piles up in the disk to make it unstable (Kenyon et al.
1990, 1994).

Note also that an ``external region'' is not necessarily physically
unconnected with the Class 0 source; i.e., this material may
eventually collapse onto and be incorporated into the central star(s).
Whether this is necessarily the case, however, is not clear.

\subsection{Class 0 vs. Class I: the evolutionary model vs. initial density}

AWB suggested that the Class 0 sources represent the earliest
phase of protostellar collapse, based on the large amount of circumstellar
material in the near environment (compared with the likely
mass of the central protostellar object). 
Henriksen, Andr\'e, \& Bontemps (1997) elucidated a 
picture which in principle provides a theoretical basis for this
evolutionary scheme.

Henriksen et al. pointed out that theoretical models of clouds 
initially in hydrostatic equilibrium, such as the isothermal
Bonnor-Ebert sphere (e.g., Foster \& Chevalier 1993), 
tend to have flat inner density distributions surrounded by an outer envelope
of roughly $\rho \propto r^{-2}$.  In addition, Henriksen et al. emphasized
the observational evidence for flattened central regions in pre-stellar
cores mentioned in \S 4.1. 
When the initially flat inner regions of such a cloud fall in, 
high mass infall rates are produced (see also Foster \& Chevalier 1993);
material from the outer envelope falls in later at slower rates.
This leads to time-dependent infall rates, with the highest rates
occuring during the earliest phases of collapse, identified as the Class 0
phase.
 
This picture, however, does not explain why Class 0
sources tend to be found in dense star forming regions.
Moreover, while the general picture of decreasing infall rate with
increasing age is quite plausible, it is not clear whether such a
Class 0 phase would dominate protostellar infall, because
the amount of mass in the flattened inner portions of the
initial cloud is generally smaller than the mass in
the envelope, implying that most of the protostellar mass must
be added during the envelope infall (Class I) phase.  Henriksen et al. (1997)
show that, for a flat region of a cloud to contain half of the total
cloud mass, the $\rho \propto r^{-2}$ envelope can only extend out
a factor of 1.6 in radius beyond the flat region --  essentially, the cloud
is almost entirely at constant density, with only a small outer
envelope region, inconsistent with observations of typical pre-stellar cores.
Henriksen et al. suggest that clouds with steeper edges,
$\rho \propto r^{-3}$, could alleviate this problem while maintaining
the required large factor $\sim 10$ between Class I and Class 0 lifetimes;
but they do not show that the very slow infall implied by these
models for the Class I phase is consistent with SED models.

Alternatively, one might suppose that
the high environmental densities characterizing the birthsites of
Class 0 sources {\it guarantee that
envelope masses will be very large on small scales.}
In other words, the Class 0 sources might be the Class I
sources of dense star-forming regions.

As a specific example, consider two protostellar clouds of the same mass but 
formed in regions of differing average density.
The characteristic infall rates will scale roughly as the cloud mass
divided by the free fall time, which in turn scales roughly as the inverse
square root of the mean density; thus, $\md \propto  M/<\rho>^{-1/2} 
\propto <\rho>^{1/2} \propto R^{-3/2}$, where $R$ is the outer radius of the 
cloud. (Let us fix the central star mass in order to discuss the difference
in infall rates.) If we assume that the Class 0 source has 
an infall rate of $10^{-4} \msunyr$, and a Class I source has an infall rate of
$4 \times 10^{-6} \msunyr$, then if the two protostellar clouds originally
had the same mass, the factor of 25 in $\md$ implies that the Class I source
would have to be larger in radius by a factor of 8.5.  In other words, the 
mass within a 2000 AU (12 arcsec at 160 pc) radius in a Class 0 
source should be compared to the mass within a radius of 16,000 AU ($\sim~100$~arcsec) 
in a Class I source to properly assess the relative evolutionary states of these objects.

The molecular surveys of Myers and collaborators
(e.g., Benson \& Myers 1989; Zhou et al. 1989) have indicated that many cores, such
as those in Taurus, have substantial spatial extensions.  
For instance, even the lowest-mass ammonia cores in the
survey of Benson \& Myers (1989) have $0.8 - 0.6 \msun$
of material, comparable to that of the Class 0 sources, 
but distributed within structures of (half-power diameters) 
$\sim 10,000$~AU $\sim 0.05$~pc
rather than 2000~AU~$= 0.01$~pc.  In many other cases the ammonia core
mass is considerably larger than $1 \msun$ and spread over larger scales.
Whether a substantial portion of the material in a typical Taurus 
molecular cloud core actually is incorporated into a particular protostellar 
source is not clear --or whether it makes only one star.  However, the same
could be said of Class 0 sources.  Neglect of the possible importance
of extended material could bias the comparison between the two types of
sources. 

Bontemps et al. (1996) attempted a systematic assessment of Class 0 and 
Class I source masses from measurements of 1.3mm continuum emission.
In computing dust masses they adopted a ``typical'' source diameter of 1 arcmin
over which to integrate the emission.  In doing so, Bontemps et al.
recognized that they might be measuring only the ``inner'' envelope
mass of ``isolated'' protostars such as those in Taurus.  Moreover,  
if our suggestion of the presence of significant ``external'' material 
in VLA 1623 is correct, then Class 0 mass estimates might be biased
by the inclusion of mass that may not end up on the protostar,
especially if integrated over a radius of 30 arcsec $\sim 4800$~AU,
rather than the 2000 AU radius adopted in the previous sections.
Given the large differences in densities observed in
star-forming regions, the systematic biases in mass estimates
from adopting uniform source sizes can be of an order of magnitude,
which could easily affect assessments of the relative evolutionary
states of Class 0 and Class I sources.  Unfortunately, the observational
problem of deciding where a protostellar cloud ``stops'' is likely
to be quite difficult.  Yet, without consideration of this potential bias,
it seems premature to conclude that all Class 0 sources must be in
a radically different phase of evolution than Class I sources.

In Taurus, the source statistics of Class I sources suggest a lifetime
of $1 - 2 \times 10^5$~yr (Kenyon et al. 1990, 1994).  
If one adopts the infall rates of $\sim 4 \times 10^{-6} \msunyr$
typically estimated from SED fitting (ALS; KCH), then one would
expect that approximately $0.4 - 0.8 \msun$ is added to protostars
during the Class I phase - which is comparable to the {\it total}
mass accreted onto the average Taurus T Tauri star.  
It is difficult to see why the Class I period
is not an important part of the protostellar phase
in Taurus, unless somehow
infall rates have been overestimated by an order of magnitude.
If, on the other hand, Class I sources {\it are} the protostars
of low-density regions like Taurus, then there is no discrepancy
to explain.  The two borderline Class 0
objects in Taurus (L1551 IRS 5 and L1527; Chen et al. 1995; or L1527
and IRAM 04191+1522; Andr\'e et al. 1999) could be
among the youngest protostellar sources in Taurus (Ladd et al. 1998);
however, they might also be objects formed from particularly
dense subregions within Taurus.

To summarize, we agree with AWB and AWB2 that the Class 0 sources are very
young in an absolute sense because their envelopes must collapse so rapidly.
However, we suggest that the high infall rates of Class 0 sources can 
result also from the high densities of their initial pre-stellar cloud 
cores rather than from always being the very first phases of collapse.  
Whether Class 0 sources are systematically
younger than the Class I sources in a relative sense,
i.e. have a much smaller fraction of their total mass accumulated into
a central protostar, is far from clear given the observational biases
and uncertainties in defining envelope limits.
Some Class 0 sources may also be young in this relative sense, but
it is difficult to support the conclusion that they are all
the ``youngest protostars''.

\subsection{Dense environments and the standard picture}

Infall rates on the order of 
$10^{-4}\msun \mathrm{yr^{-1}}$ seem to be needed to account for Class 0
sources. This infall rate is much higher 
than the typical infall rates $\sim 4 \times 10^{-6}\msun \mathrm{yr^{-1}}$
for Class I sources in Taurus-Auriga (KCH).
In the standard singular isothermal sphere model, the infall rate
is $\md \sim c_s^3 /G$; thus the mean Taurus infall rate requires a
sound speed $c_s \sim 0.25 \kms$ or a gas temperature of $18$~K assuming
a mean molecular weight of 2.3,  only slightly larger than
typical temperatures found in the region.  In contrast,
to obtain infall rates of $\sim 10^{-4}\msun \mathrm{yr^{-1}}$ in a cloud
supported only by thermal gas pressure would require unrealistically
high cloud temperatures of about 150K.  Thus, some other source of 
support must be involved if the cloud is initially in hydrostatic equilibrium.

These considerations emphasize the importance of the density of the
protostellar environment to an understanding of Class 0 sources.  The related 
question is how do such dense regions of material arise? 
If such regions are in hydrostatic equilibrium, 
turbulent motions undoubtedly play a role in supporting denser clouds
(e.g., Lizano \& Shu 1989; Fuller \& Myers 1993).  
On the other hand, since the sources of turbulence
are presumably non-isotropic and inhomogeneously distributed (i.e., winds,
magnetic waves excited by moving dense gas clumps, etc.)
arranging static situations seems very difficult.  In addition, supersonic
magnetic turbulence may decay rapidly (Stone, Ostriker, \& Gammie 1998).
An alternative view is that
dense streams of gas in the ISM, however produced, collide and dissipate energy
in a shock (e.g., Ballesteros-Paredes, V\'azquez-Semadeni \& Scalo 1999), 
resulting in dense regions which
then undergo gravitational collapse.  In such a general picture, there is no
obvious reason why densities of protostellar regions need be restricted in
any way to levels supportable by thermal gas pressure.  In this way protostars
with dense envelopes and rapid infall rates could be produced without reference
to the usual constraints of thermal support (ALS).  The Class 0 sources, or at
least some of them, could be telling us more about protostellar formation in extreme
environments than about the very earliest phases of protostellar collapse.

A related question is, why did the standard model of singular isothermal sphere
collapse work so well in Taurus?  We suggest that this is simply due to
initial conditions.  In the absence of any complicating factors --magnetic
field support, turbulent motions, etc. -- thermal gas pressure provides
an absolute lower limit to the forces resisting gravitational collapse,
especially as cosmic ray heating seems to maintain cloud temperatures
at a minimum value of $\sim 10$~K in the absence of any other heat sources
(Goldsmith 1988).  Absent substantial turbulent support, or high-density flows,
Taurus may only represent a limiting case of star formation.

\section{Summary}
We have described radiative transfer calculations for the spectral energy
distributions of recently identified Class 0 protostars. Our models require
infall rates on the order of $10^{-4} \mathrm{\msun yr^{-1}}$ to account 
for the 
observed SEDs. The sub-millimeter intensity profiles of the Class 0 source 
VLA 1623 apparently imply a shallower density distribution than that 
produced by infall. But we show that material exterior to the infall region 
may contribute to the observed intensity profiles and thus infall models 
(with exterior emission) can be consistent with observations. The high
infall rates in Class 0s, as required by our models and implied by a variety
of observations, may be inevitable given their dense environments. Thus, 
we suggest that protostars in dense regions are more likely to appear as
Class 0 sources while those in lower-density regions manifest Class I 
characteristics. The Class 0/Class I division may thus represent initial
conditions more than evolutionary status, at least for some sources. 

\bigskip
We are grateful to the referee, Dr. S. Terebey, for a thorough review.
This work was supported by the Smithsonian Institution.

\newpage

\centerline{\bf Figure Captions}
Figure 1 - (a) Spectral energy distribution of VLA 1623 using data from
AWB. (The expected contribution from free-free emission at long wavelengths
has been subtracted.) The solid curve is an infall model with $\Md
\sim 2 \times 10^{-4} \msun$, and the dotted curve is a model with
$\Md \sim 10^{-5} \msun$. (b) Spectral energy distribution of
HH24MMS using data from Ward-Thompson et al. (1995). The solid curve is
an infall model with $\Md \sim 3 \times 10^{-4} \msun$.

Figure 2 - Comparison of the radial intensity profiles of VLA 1623 at
800$\mu$m, 450$\mu$m, and 350$\mu$m with the basic infall model in which 
$\rho \propto r^{-3/2}$ ({\it dotted line}), the infall model where 
$T_{min}$=20K ({\it dashed line}), and the composite model described in 
Section 3 which includes a contribution from regions exterior to the 
infalling core ({\it solid line}). The model output was convolved with
the appropriate beam size at each wavelength and scaled to a distance
of 160 pc (13.5''=2160AU, 7.5''=1200AU, and 6.3''=1000AU, respectively).

Figure 3 - Comparison of the spectral energy distribution of VLA 1623 to
the best-fit composite model ({\it solid line}). The dotted line represents
emission of the infall region while the dashed line shows the
contribution of the exterior ``slab.''

\clearpage
\begin{figure}
\epsscale{1.0}
\plotone{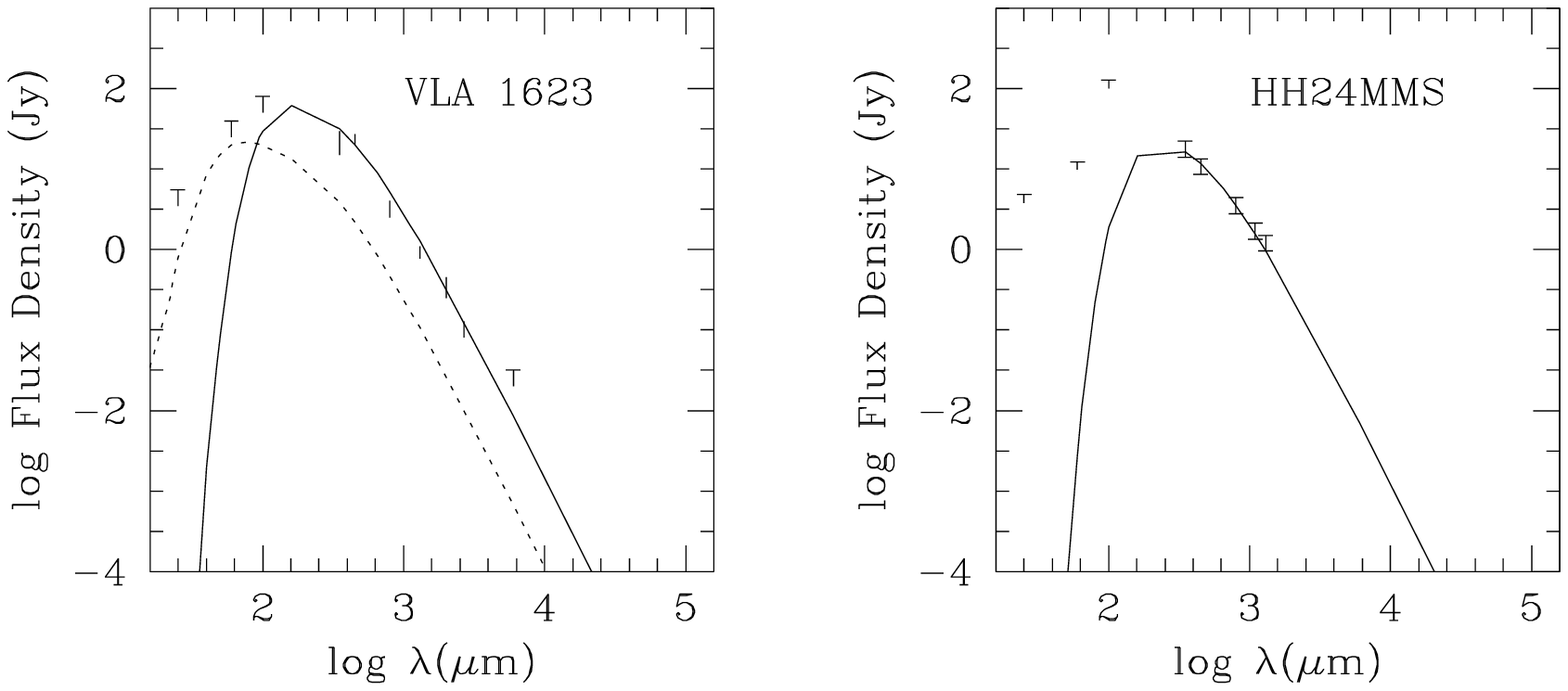}
\end{figure}

\clearpage
\begin{figure}
\epsscale{1.0}
\plotone{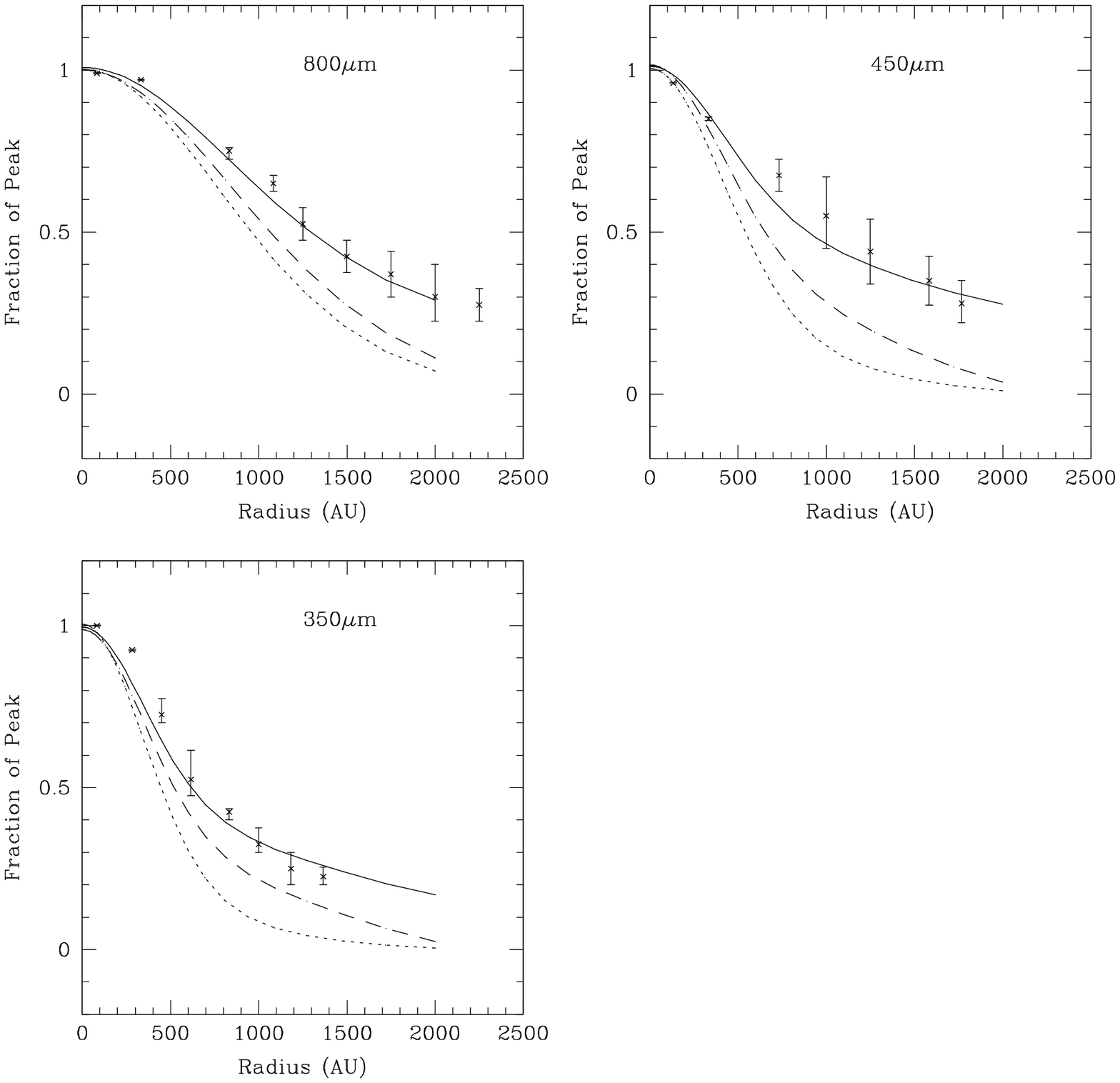}
\end{figure}

\clearpage
\begin{figure}
\epsscale{1.0}
\plotone{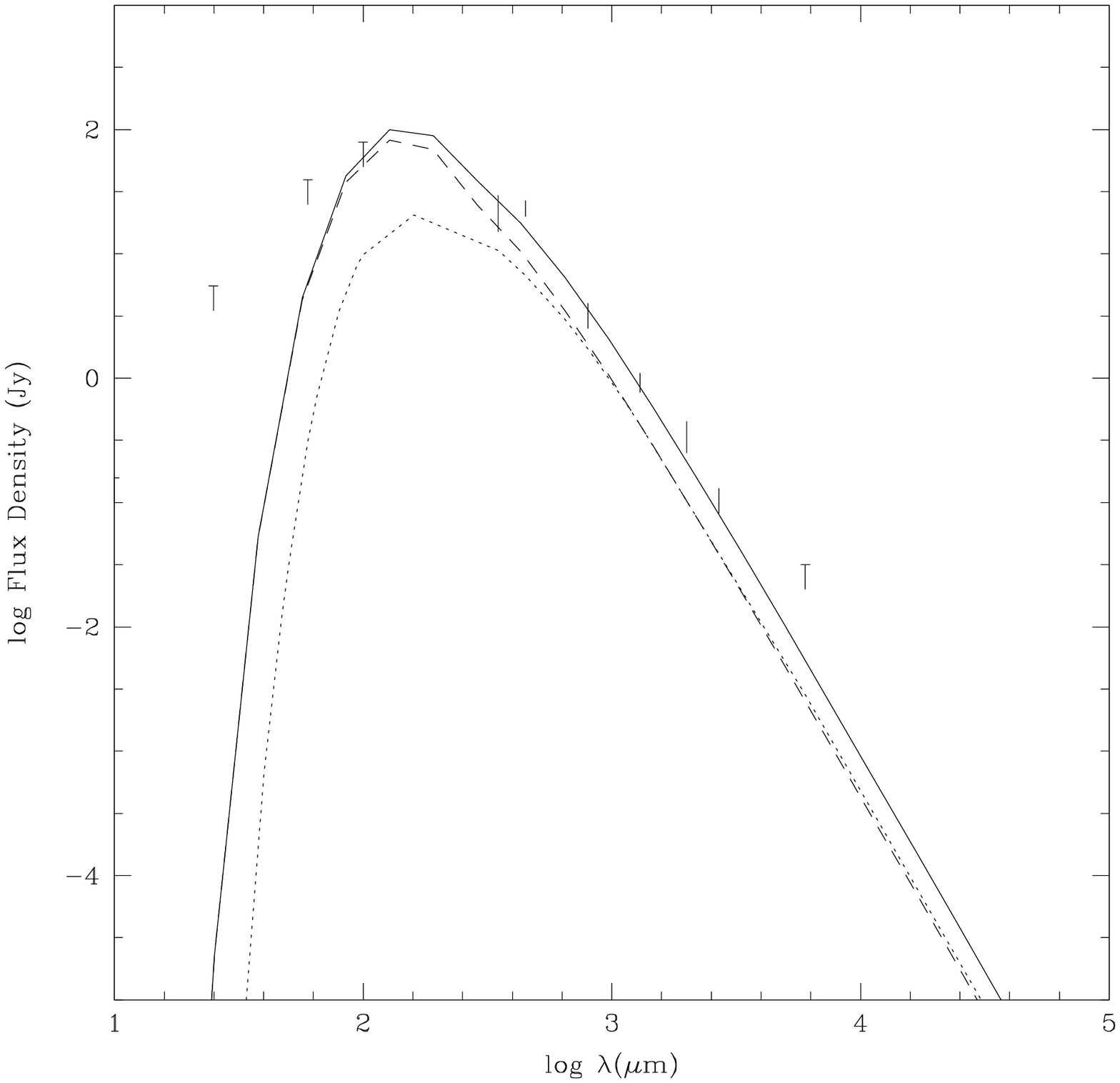}
\end{figure}


\newpage
\begin{references}
\reference{} Adams, F.C., Lada, C.J., \& Shu, F.H. 1987, ApJ, 312, 788 (ALS)
\reference{} Adams, F.C. \& Shu, F.H. 1986, ApJ, 308, 836
\reference{} Andr$\acute{\rm e}$, P., Martin-Pintado, J. Despois, D., \& Montmerle, T. 1990a, A\&A, 236, 180
\reference{} Andr$\acute{\rm e}$, P., Montmerle, T., Feigelson, E.D., \& Steppe, H. 1990b, A\&A, 240, 321
\reference{} Andr$\acute{\rm e}$, P. 1997, in \emph{Herbig-Haro Flows and the Birth of Low Mass Stars}, ed. B. Reipurth \& C. Bertout (Dordrecht: Kluwer), 483
\reference{} Andr$\acute{\rm e}$, P., \& Montmerle, T. 1994, ApJ, 420, 837
\reference{} Andr$\acute{\rm e}$, P., Motte, F., \& Bacmann, A. 1999, ApJ, 513, L57
\reference{} Andr$\acute{\rm e}$, P., Ward-Thompson, D. \& Barsony, M. 1993, ApJ, 406, 122 (AWB)
\reference{} Andr$\acute{\rm e}$, P., Ward-Thompson, D. \& Barsony, M. 2000, in Protostars and Planets IV,
eds. V. Mannings, A.P. Boss, \& S.S. Russell (Tucson: Univ. of Arizona Press), 59 (AWB2)
\reference{} Andr$\acute{\rm e}$, P., Ward-Thompson, D., \& Motte, F. 1996, A\&A, 314, 625
\reference{} Ballesteros-Paredes, J. V\'azquez-Semadeni, E.,  \& Scalo, J.M. 1999, ApJ, 515, 286 
\reference{} Barsony, M. 1994, in \emph{Clouds, Cores and Low Mass Stars}, ed. D.P. Clemens \& R. Barvainis (Provo: Astronomical Society of the Pacific), 197
\reference{} Benson, P.J. \& Myers, P.C. 1989, ApJS, 71, 89
\reference{} Bontemps, S., Andr$\acute{\rm e}$, P., \& Ward-Thompson, D. 1995, A\&A, 297, 98
\reference{} Bontemps, S., Andr$\acute{\rm e}$, P., Terebey, S., \& Cabrit, S. 1996, A\&A, 311, 858
\reference{} Calvet, N., Hartmann, L., Kenyon, S.J. \& Whitney, B.A. 1994, ApJ, 434, 330
\reference{} Chandler, C.J., \& Richer, J.S. 2000, ApJ, 530, 851
\reference{} Chen, H., Myers, P.C., Ladd, E.F., \& Wood, D.O.S. 1995, ApJ, 445, 377
\reference{} Chini, R., Kr\"{u}gel, E., Haslam, C.G.T., Kreysa, E., Lemke, R., Reipurth, B., Sievers, A., \& Ward-Thompson, D. 1993, A\&A, 272, L5
\reference{} Draine, B.T. \& Lee, H.M. 1984, ApJ, 285, 89
\reference{} Ferreira, J. \& Pelletier, G. 1995, A\&A, 295, 807
\reference{} Foster, P.N. \& Chevalier, R.A. 1993, ApJ, 416, 303
\reference{} Fuller, G.A. \& Myers, P.C. 1993, ApJ, 418, 273
\reference{} Galli, D. \& Shu, F.H. 1993a, ApJ, 417, 220
\reference{} Galli, D. \& Shu, F.H. 1993b, ApJ, 417, 243
\reference{} Goldsmith, P.F. 1988, in \emph{Molecular clouds in the Milky Way}, Lecture Notes in Physics, ed. R.L. Dickman, R.L. Snell, J.S. Young (Springer-Verlag), vol. 315, 1
\reference{} Hartmann, L., Boss, A., Calvet, N., \& Whitney, B. 1994, ApJ, 430, 49
\reference{} Hartmann, L., Calvet, N., \& Boss, A. 1996, ApJ, 464, 387
\reference{} Henriksen, R., Andr\'e, P., \& Bontemps, S. 1997, A\&A, 323, 549
\reference{} Hogerheijde, M.R., \& Sandell, G. 2000, ApJ, 534, 880
\reference{} Hurt, R.L. \& Barsony, M. 1996, ApJL, 460, L45
\reference{} Kenyon, S.J., Hartmann, L.W., Strom, K.M., \& Strom, S.E. 1990, AJ, 99, 869
\reference{} Kenyon, S.J., Calvet, N. \& Hartmann, L. 1993, ApJ, 414, 676 (KCH)
\reference{} Kenyon, S.J., Gomez, M., Marzke, R.O., \& Hartmann, L. 1994, AJ, 108, 251
\reference{} Lada, C.J. 1987, in \emph{Star Forming Regions}, ed. M. Peimbert \& J. Jugaku (Dordrecht: Reidel), 1
\reference{} Lada, C.J. \& Wilking, B. A. 1984, ApJ, 287, 610
\reference{} Ladd, E.F., Fuller, G.A., \& Deane, J.R. 1998, ApJ, 495, 871
\reference{} Larson, R.B. 1969, MNRAS, 145, 271
\reference{} Lefloch, B., Eisenhoffel, J., \& Lazareff, B. 1996, A\&A, 313, 17
\reference{} Leous, J.A., Feigelson, E.D., Andr$\acute{\rm e}$, P., \& Montmerle, T. 1991, ApJ, 379, 683
\reference{} Li, Z.Y. \& Shu, F.H. 1996, ApJ, 472, 211
\reference{} Lizano, S., \& Shu, F.H. 1989, ApJ, 342, 834
\reference{} Mardones, D., Myers, P.C., Tafalla, M., Wilner, D.J., Bachiller, R., \& Garay, G. 1997, ApJ, 489, 719
\reference{} Motte, F., Andr$\acute{\rm e}$, P., \& Neri, R. 1998, A\&A, 336, 150
\reference{} Mouschovias, T.Ch. 1991, ApJ, 373, 169
\reference{} O'Linger, J.C., Wolf-Chase, G., \& Barsony, M. 1997, BAAS, 191, 716
\reference{} Penston, M.V. 1969, MNRAS, 144, 425
\reference{} Pudritz, R.E., Wilson, C.D., Carlstrom, J.E., Lay, O.P., Hills, R.E., Ward-Thompson, D. 1996, ApJL, 470, L123
\reference{} Shu, F.H. 1977, ApJ, 214, 488
\reference{} Shu, F.H., Adams, F.C., \& Lizano, S. 1987, ARA\&A, 25, 23
\reference{} Shu, F., Najita, J., Ostriker, E., Wilkin, F., Ruden, S., \& Lizano, S. 1994, ApJ, 429, 781
\reference{} Stone, J.M., Ostriker, E.C., \& Gammie, C.F. 1998, ApJ, 508, L99
\reference{} Terebey, S., Chandler, C.J., \& Andr\'e, P. 1993, ApJ, 414, 759
\reference{} Terebey, S., Shu, F.H., \& Cassen, P. 1984, 286, 529 (TSC)
\reference{} Ward-Thompson, D., Chini, R., Kr\"{u}gel, E., Andr$\acute{\rm e}$, P., \& Bontemps, S. 1995, MNRAS, 274, 1219
\reference{} Ward-Thompson, D., Motte, F., \& Andr\'e, P. 1999, MNRAS, 305, 143
\reference{} Ward-Thompson, D., Scott, P.F., Hills, R.E., \& Andr\'e, P. 1994, MNRAS, 268, 276
\reference{} Zhou, S., Wu, Y., Evans, N.J.,II, Fuller, G.A., \& Myers, P.C. 1989, ApJ, 346, 168
\end{references}
\end{document}